\documentclass{article}

\usepackage{microtype}
\usepackage{graphicx}
\usepackage{subcaption}
\usepackage{booktabs}
\usepackage{kbordermatrix}
\usepackage{tikz}
\usepackage{chemfig}
\setchemfig{atom sep=2em}
\usepackage{mod}
\usepackage[colorlinks=true, allcolors=Blue]{hyperref}
\hypersetup{
    colorlinks,
    linkcolor = Blue,
    citecolor = blue,
    urlcolor = BlueViolet
}

\usepackage[accepted]{icml2025}
\usepackage{amsmath}
\usepackage{amssymb}
\usepackage{mathtools}
\usepackage{amsthm}

\usepackage[capitalize,noabbrev]{cleveref}

\theoremstyle{plain}

\theoremstyle{definition}

\theoremstyle{remark}

\usepackage[textsize=tiny]{todonotes}

\icmltitlerunning{MILP to search for pathways in Reaction Networks}

\begin{document}

\twocolumn[
\icmltitle{\textsc{Finding Thermodynamically Favorable Pathways\\ in Chemical Reaction Networks using Flows in Hypergraphs\\ and Mixed Integer Linear Programming}}

\icmlsetsymbol{equal}{*}

\begin{icmlauthorlist}
\icmlauthor{Adittya Pal}{sdu,equal}
\icmlauthor{Rolf Fagerberg}{sdu}
\icmlauthor{Jakob Lykke Andersen}{sdu}
\icmlauthor{Christoph Flamm}{uvie}
\icmlauthor{Peter Dittrich}{fsu}
\icmlauthor{Daniel Merkle}{sdu,bu}
\end{icmlauthorlist}

\icmlaffiliation{sdu}{Department of Mathematics and Computer Science, University of Southern Denmark, Campusvej 55, 5230 Odense M, Denmark\\}
\icmlaffiliation{uvie}{Department of Theoretical Chemistry, University of Vienna, W\"{a}hringer Stra{\ss}e 17, 1090 Wien, Austria\\}
\icmlaffiliation{fsu}{Department of Mathematics and Computer Science, Friedrich Schiller University Jena, F{\"u}rstengraben, 07743, Jena, Germany\\}
\icmlaffiliation{bu}{Faculty of Technology, Bielefeld University, Postfach 10 01 31, 33501 Bielefeld, Germany}

\icmlcorrespondingauthor{Adittya Pal}{adpal@imada.sdu.dk}

\icmlkeywords{thermodynamics of chemical reaction networks, mixed integer linear program, pathways in chemical reaction networks, linear optimization}

\vskip 0.3in
]

\printAffiliationsAndNotice{\icmlEqualContribution} 

\begin{abstract}
The search for pathways that optimize the formation of a particular target molecule in a reaction network is a key problem in many settings, including reactor systems.
Chemical reaction networks are mathematically well represented as hypergraphs, modeling that facilitates the search for pathways by computational means. 
We propose to enrich an existing search method for pathways by including thermodynamic principles. In more detail, we give a mixed-integer linear programming (mixed ILP) formulation of the search problem into which
we integrate chemical potentials and concentrations for individual molecules, enabling us to constrain the search to return pathways containing only thermodynamically favorable reactions.
Moreover, if multiple possible pathways are found, we can rank these by objective functions based on thermodynamics.
As an example of use, we apply the framework to a reaction network representing the HCN-formamide chemistry. Alternative pathways to the one currently hypothesized in the literature are queried and enumerated, including some that score better according to our chosen objective function.
\end{abstract}

\section{Introduction}\label{sec: introduction}

Equilibrium thermodynamics has been often used to determine the spontaneity of an isolated chemical reaction. Specifically, from the Gibbs free energy one can find the driving forces behind a reaction and determine whether its occurrence is favorable or not. However, in practical scenarios, reactions are rarely encountered in isolation. Rather, they are components of larger interconnected networks, where the products of one reaction may be the reactants of numerous other reactions. Such networks are called \emph{reaction networks}.
In this paper, an existing computational methodology for searching for pathways in reaction networks via flow queries is extended with the ability to include equilibrium thermodynamics among the search constraints.

\subsection{Context and Motivation}

Chemical reaction networks are recurrent in real-world scenarios: cellular metabolic processes,
interactions among atmospheric constituents, environmental phenomena such as cycling of nutrients and pollutants,
the human gut microbiome, and industrial chemical reactors.
Modeling these networks is challenging due to the inherent complexities of such systems.
One type of complexity arises from nonlinear dynamics, which appear in particular when transcending one-to-one reactions.
Another type of complexity arises from the existence of parallel and alternative sequences of reactions,
termed pathways, which may give the same overall effect but involve distinct yet intersecting sets of molecules.
Such pathways in reaction networks are manifested in diverse contexts:
metabolic pathways in organisms for cellular metabolic networks, nutrient cycles in the environment, and synthesis plans in chemical reaction networks.

The exploration of reaction networks to synthesize molecules has primarily been done experimentally in the wet laboratory. However, this process is labor intensive, costly, and limits the pace of exploration. \emph{In silico} modeling and exploration of reaction networks have the potential to speed up the process and guide manual efforts. The finer the physical details included in the modeling of the reaction networks, the more accurate the results can be expected to be. The ultimate way of doing that would be to construct the complex free energy landscape for the full reaction network, which then, in principle, would allow for the full simulation of reactions, as these are ultimately guided by the laws of thermodynamics.
However, we believe that this is currently past any modeling and computational capacities available. A first and more realistic step towards the goal is using chemical potentials to determine if a given reaction is favorable in thermodynamical terms, and then use only the favorable reactions in the exploration of the reaction networks.

\subsection{Previous Work}

Thermochemistry data have been used to suggest pathways in reaction networks through pathfinding algorithms and a linear combination of the lowest cost pathways in that network~\cite{synthesis_planning_CRN_thermodynamics}. 
In thermodynamics-based metabolic flux analysis~\cite{hatz_thermo_flow}, the mass balance constraints of metabolic flux analysis are augmented by additional thermodynamic constraints to produce flux distributions and metabolite activity profiles free from thermodynamic infeasibilities.
In \cite{seifert_thermodynamics_CRN_JCP}, nonequilibrium thermodynamics was developed to explore reaction networks, employing a stochastic trajectory-based approach to define energy, entropy, and heat dissipation along reactions.
Building upon stochastic thermodynamics, a nonequilibrium thermodynamic framework for open reaction networks has been formulated~\cite{esposito_thermodynamics_CRN_PRX}, governed by deterministic rate equations with mass action kinetics and introducing a nonequilibrium Gibbs free energy quantifying the minimal work required to induce nonequilibrium states from an equilibrium state.
This nonequilibrium Gibbs free energy is minimized as the open detailed-balanced network relaxes to an equilibrium state, aligning with the maximum entropy principle.
Circuit theory has been applied to reaction networks~\cite{esposito_circuit_theory_CRN_PRX} extending the chemical analogs of Kirchhoff’s laws to predict reaction currents within complex reaction networks.
\cite{ILP_paper2} utilizes linear programming to identify complex balanced realizations of mass action kinetics in reaction networks.
\cite{ILP_paper2, ILP_paper3, ILP_paper4} have introduced numerical procedures for determining weakly reversible reaction networks using integer linear programming (ILP), emphasizing the importance of ILPs in identifying specific network properties.
Mixed integer linear programming has been employed to prune reaction networks using experimental data and to understand complex chemical systems in~\cite{ilp_CRN_infer}.

Much of the work cited in the preceding paragraph primarily focuses on theoretical formulations, with few executable implementations.
In contrast, MØD~\cite{mod_web,jakob_software} is a software package developed to generate and analyze reaction networks.
M{\O}D models molecules using undirected graphs with attributes, hence with each atom explicitly represented, and reactions are modeled by rules for transforming molecule graphs.
Reaction networks can automatically be generated from a set of rules and a set of starting molecules using the graph transformation engine of M{\O}D.
The outcome is a directed multihypergraph representing the reaction network.
The pathways in such reaction networks can then be queried using integer hyperflows as a formal model~\cite{jakob_integer_hyperflows}
and the reaction network can be queried using computational means for the presence of the specified pathway.
This framework allows for a flexible, yet precise specification of pathways and a search for such pathways by computational means.
However, the methodology~\cite{jakob_integer_hyperflows} currently does not incorporate any dynamics in the modeling of the reaction network.

\subsection{Our Contribution}

In this paper, we incorporate thermodynamic principles into the existing hyperflow model in M{\O}D.
In particular, we develop methods based on mixed integer linear programming (MILP) that allow thermodynamic constraints to be added to the specification of pathways when searching for these in the M{\O}D framework.
Our methods can return several alternative pathways, and these can be ranked on the basis of thermodynamic-based metrics. 

We note that returning multiple pathways can significantly increase the practical value of the investigations.
Models of reaction networks and estimations of thermodynamic parameters are inherent simplifications of reality.
Hence, any single pathway returned may turn out to be less desirable in the face of further details of the chemical situation.
Hence, having a number of highly ranked pathways available from which the practitioner can choose is a more robust strategy.

In our method, we need to assign chemical potentials to the molecules in the reaction network to calculate the Gibbs free energies for its reactions.
In the current implementation, we use the semi-empirical method GFN2-xTB~\cite{grimme_xtb},
which provides a good balance between accuracy of the thermodynamic parameters and the ability to handle larger molecules.
However, it can be replaced by any other method to assign chemical potentials if the user desires so.
For generality we refer to the chosen method as a thermodynamic oracle.
We note that using a thermodynamic oracle to calculate the chemical potentials of the molecules in an automated fashion has the added benefit of being applicable in rule-based generative systems where the chemical space is expanded on the fly and the molecules generated are unknown at the start.

To the best of our knowledge, there currently does not exist a computational method for searching for pathways in reaction networks based on thermodynamic constraints that combines MILP with the wide applicability and speed of semiempirical tight-binding methods.
In Section \ref{sec: discussions}, we give further details of the comparison of our work with existing frameworks.

\subsection{Paper Structure}

The remainder of this paper is organized into the following sections.
In Section \ref{sec: definitions}, we formulate our modeling of reaction networks and pathways.
In Section \ref{sec: methods}, which constitutes the core part of the paper, we give the details of our approach.
In Section \ref{sec: results}, we demonstrate the usability of our method, by applying them on a reaction network involving hydrogen cyanide~(\chemfig{HCN}).
In Section \ref{sec: discussions}, we give the comparison mentioned above with some existing frameworks.
In conclusion, we summarize the future directions of the work as a follow-up to the present.
Some additional technical details and discussions are given in the Appendices \ref{appendix}.

\section{Definitions}\label{sec: definitions}

In this section, we present our model for reaction networks and pathways,
namely directed hypergraphs and integer hyperflows, respectively.
This section is based on \cite{jakob_integer_hyperflows} which have used these concepts.

\subsection{Directed Hypergraphs}\label{sec:hypergraphs}

\begin{figure*}[ht]
    \centering
    \vspace{0.5cm}
    \begin{subfigure}{0.5\textwidth}
        \includegraphics[width=\textwidth]{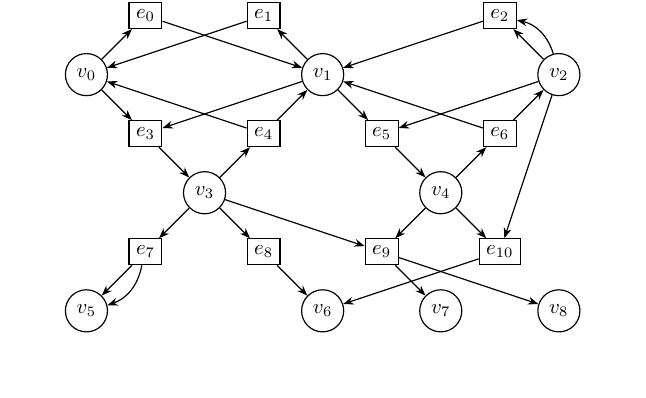}
        \caption{\label{fig: hypergraph1}}
    \end{subfigure}%
    ~
    \begin{subfigure}{0.5\textwidth}
        \includegraphics[width=\textwidth]{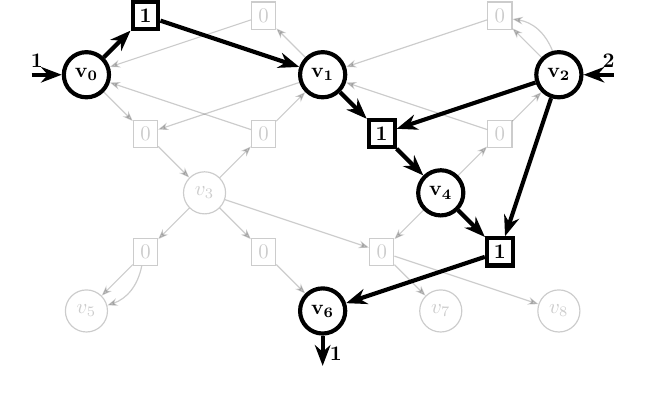}
        \caption{\label{fig: hyperpath1}}
    \end{subfigure}
    \begin{subfigure}{0.5\textwidth}
        \includegraphics[width=\textwidth]{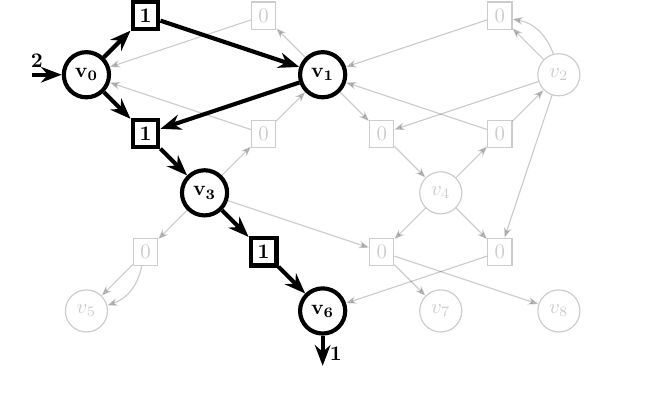}
        \caption{\label{fig: hyperpath2}}
    \end{subfigure}%
    ~
    \begin{subfigure}{0.5\textwidth}
        \includegraphics[width=\textwidth]{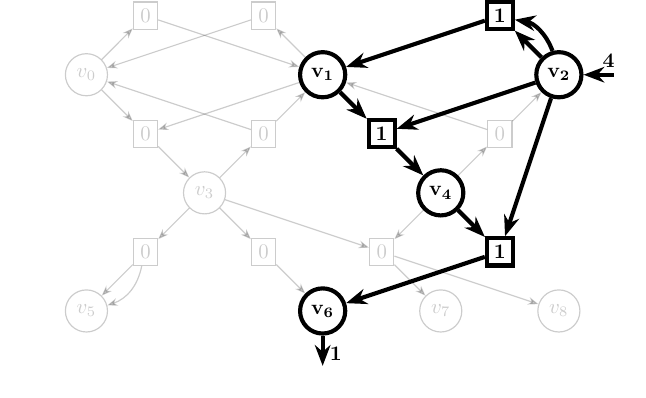}
        \caption{\label{fig: hyperpath3}}
    \end{subfigure}    
    \caption{A directed hypergraph \ref{fig: hypergraph1}, where circles are vertices and the squares are hyperedges.
    The other three subfigures \ref{fig: hyperpath1}, \ref{fig: hyperpath2}, and \ref{fig: hyperpath3} show possible integer hyperflows in the hypergraph, each representing a pathway,
    where hyperedges with non-zero flow are drawn in bold.
    The inflow and the outflow are depicted as arrows in and out of the source and target vertices of the pathway.
    Hyperedges with multiple copies of a vertex as source or target are depicted with parallel arrows.
    See for instance~$e_7$, which has $v_5$ twice as a target and thus represents the reaction $v_3\rightarrow 2 v_5$.}
    \label{fig: hypergraph}
\end{figure*}

A reaction network consists of a set of molecules $V$ and a set of reactions $E$.
Each reaction $e\in E$ is generally a many-to-many mapping of molecules,
which can be written as the formal sum:
\begin{equation*}
        \sum_{v\in V} s_{ve}^{-} v \longrightarrow \sum_{v\in V} s_{ve}^{+} v 
        \label{chemical_equation}
\end{equation*}
where $s_{ve}^{-}$ and $s_{ve}^{+}$ denote the stoichiometric coefficients, that is, the number of molecules of $v$ that are used, respectively produced, by reaction~$e$.
The sum is over the entire set of molecules $V$ in the reaction network,
and for molecules $v$ that do not participate in a particular reaction $e$, we have $s^-_{ve} = s^+_{ve} = 0$.
Hence, a reaction network is formally well represented as a directed multi-hypergraph~$(V, E)$,
where the set~$V$ of vertices represents molecules and the set~$E$ of directed multihyperedges represents reactions.
A directed multi-hyperedge $e\in E$ is an ordered pair $(e^-, e^+)$ of multisets of vertices.
The elements of $e^-$ model the reactants of the reaction, with each reactant $v$ appearing $s^-_{ve}$ times,
and correspondingly $e^+$ and $s^+_{ve}$ model the products of the reaction
(note that the use of $^+$ and $^-$ is swapped compared to the notation in~\cite{jakob_integer_hyperflows}).
Figure~\ref{fig: hypergraph} shows a hypergraph with eight vertices and eleven hyperedges.
As examples, hyperedge~$e_0 \equiv (\{v_0\}, \{v_1\})$ represents the one-to-one reaction $v_0\rightarrow v_1$,
hyperedge~$e_3 \equiv (\{v_0, v_1\}, \{v_3\})$ represents the many-to-one reaction $v_0 + v_1\rightarrow v_3$,
hyperedge~$e_2 \equiv (\{v_2, v_2\}, \{v_1\})$ represents the many-to-one reaction $2 v_2\rightarrow v_1$,
hyperedge~$e_4 \equiv (\{v_3\}, \{v_0, v_1\})$ represent the one-to-many reaction $v_3\rightarrow v_0 + v_1$,
hyperedge~$e_7 \equiv (\{v_3\}, \{v_5, v_5\})$ represent the one-to-many reaction $v_3\rightarrow 2 v_5$, and
hyperedge~$e_9 \equiv (\{v_3, v_4\}, \{v_7, v_8\})$ represents the many-to-many reaction $v_3 + v_4\rightarrow v_7 + v_8$.

On a related note, standard usage of the term `reaction' implies the breaking and forming of chemical bonds.
However, physical processes that do not alter the set of bonds in a molecule yet involve energy transformations, thereby affecting the molecule's ability to participate in a reaction.
For instance, some reactions occur only when a molecule is excited into a higher energy state, or certain molecules (such as phosphorus or sulfur) are in specific phases.
Although such transitions are not conventionally termed `reactions,' they are elementary processes accompanied by a change in free energy.
Consequently, in the rest of this paper we will use the term \emph{reaction} to denote this broader set of relations between classes of molecules.

\subsection{Integer Hyperflows}
\label{sec:hyperflows}

A pathway is a set of reactions that transform some available source molecules into desired target molecules.
To model pathways, the concept of an \emph{integer hyperflow}~$f$ in a hypergraph~$\mathcal{H} = (V,E)$ was introduced~\cite{jakob_integer_hyperflows}, which we now explain.

A pathway is viewed as a vector $f_E \in \mathbb{N}_{0}^{|E|}$, that is, an assignment of a non-negative integer to each edge in~$E$, denoting how many times the corresponding reaction appears in the pathway.
Furthermore, two vectors $f_{\text{in}} \in \mathbb{N}_{0}^{|V|}$ and $f_{\text{out}} \in \mathbb{N}_{0}^{|V|}$ denote for each molecule in~$V$ the amount it is consumed as the source molecule and, respectively, produced as the target molecule, by the pathway.

The net effect of a pathway on the number of molecules in the system is completely specified by $f_{\text{in}}$ and $f_{\text{out}}$.
\begin{equation}
    \mathbf{S}^+\cdot f_E  - \mathbf{S}^-\cdot f_E =  f_{\text{out}} - f_{\text{in}}
    \label{int_hyperflow_matrix}
\end{equation}
where $\mathbf{S}^-$ is a $|V|\times |E|$ matrix with entries~$s_{ve}^-$  and $\mathbf{S}^+$ is a $|V|\times |E|$ matrix with entries~$s_{ve}^+$.
In hypergraph terminology, $\mathbf{S}^-$ and $\mathbf{S}^+$ are the out-incidence matrix and the in-incidence matrix, respectively.
\footnote{
    In other settings, $\mathbf{S}^-$ and $\mathbf{S}^+$ are called the reactant stoichiometric matrix and the product stoichiometric matrix.
    A single stoichiometric matrix, defined as $\mathbf{S} = \mathbf{S}^+ - \mathbf{S}^-$, is less expressive for modeling reaction networks, as the role of catalysts and autocatalytic molecules cannot be captured.
    In some contexts, $f_{\text{in}}$ and $f_{\text{out}}$ are called external compounds.
}
The matrices $\mathbf{S}^-$ and $\mathbf{S}^+$ for the hypergraph in Figure \ref{fig: hypergraph} are shown in Figure \ref{fig: stoichoimetric_matrix}.
\begin{figure*}[ht]
    \centering
    \begin{tiny}
    \[       
        \kbordermatrix{
                &e_0&e_1&e_2&e_3&e_4&e_5&e_6&e_7&e_8&e_9&e_{10}\\
            v_0 & 1 &   &   & 1 &   &   &   &   &   &   &   \\
            v_1 &   & 1 &   &   &   & 1 &   &   &   &   &   \\
            v_2 &   &   & 2 &   &   & 1 &   &   &   &   & 1 \\
            v_3 &   &   &   &   & 1 &   &   & 1 &   &   &   \\
            v_4 &   &   &   &   &   &   & 1 &   &   & 1 & 1 \\
            v_5 &   &   &   &   &   &   &   &   &   &   &   \\
            v_6 &   &   &   &   &   &   &   &   &   &   &   \\
            v_7 &   &   &   &   &   &   &   &   &   &   &   \\
            v_8 &   &   &   &   &   &   &   &   &   &   &  
        }
        \text{\hspace{1cm}}
        \kbordermatrix{
                &e_0&e_1&e_2&e_3&e_4&e_5&e_6&e_7&e_8&e_9&e_{10}\\
            v_0 &   & 1 &   &   & 1 &   &   &   &   &   &   \\
            v_1 & 1 &   & 1 &   &   &   & 1 &   &   &   &   \\
            v_2 &   &   &   &   &   &   & 1 &   &   &   &   \\
            v_3 &   &   &   & 1 &   &   &   &   &   &   &   \\
            v_4 &   &   &   &   &   & 1 &   &   &   &   &   \\
            v_5 &   &   &   &   &   &   &   & 2 &   &   &   \\
            v_6 &   &   &   &   &   &   &   &   & 1 &   & 1 \\
            v_7 &   &   &   &   &   &   &   &   &   & 1 &   \\
            v_8 &   &   &   &   &   &   &   &   &   & 1 & 
        }
    \]
    \end{tiny}
    \caption{A representation of the hypergraph from Figure \ref{fig: hypergraph} as an out-incidence matrix $\mathbf{S}^-$ (left) and an in-incidence matrix $\mathbf{S}^+$ (right).
    Only non-zero entries are shown.
    Since the hypergraph has $9$ vertices and $11$ hyperedges, the matrices have dimensions $9 \times 11$.}
    \label{fig: stoichoimetric_matrix}
\end{figure*}

To unify the three vectors~$f_E$, $f_{\text{in}}$, and $f_{\text{out}}$ into a single flow vector on hyperedges, \cite{jakob_integer_hyperflows} introduced for each $v \in V$ two so-called half-edges $e_v^+ = (\emptyset,  v)$ and $e_v^- = (v, \emptyset)$.
These represent the addition of~$v$ as source molecule to the system and the removal of~$v$ as target molecule from the system, respectively.
Setting
$$
\begin{array}{lcl}
E^+ & = & \{ e_v^+ \mid  v \in V\}\\
E^- & = & \{ e_v^- \mid  v \in V\}\\
\overline{E} & = & E \cup E^+ \cup E^-,\\
\end{array}
$$
the extended hypergraph is $\overline{\mathcal{H}} = (V, \overline{E})$
and \cite{jakob_integer_hyperflows} define an \emph{integer hyperflow} to be a vector $f \in \mathbb{N}_{0}^{|\overline{E}|}$ satisfying the flow conservation constraint
\begin{equation}
    \forall v\in V\colon \sum_{ e\in \overline{E} } s_{ve}^+ f_e - \sum_{ e\in \overline{E} } s_{ve}^- f_e = 0 
    \label{int_hyperflow_jakob}
\end{equation}
Here, $s_{ve}^-$ is an entry in the $|V|\times |\overline{E}|$ out-incidence matrix of $\overline{\mathcal{H}}$, $s_{ve}^+$ is an entry in the $|V|\times |\overline{E}|$ in-incidence matrix of $\overline{\mathcal{H}}$,
and $f_e$ is an entry in $f$.
The constraint in Equation~\eqref{int_hyperflow_jakob} is easily seen to be equivalent to the constraint in Equation~\eqref{int_hyperflow_matrix}. A detailed discussion of how integer hyperflows relate to, and differ from, other similar concepts such as elementary modes, extreme pathways, and flux balance analysis can be found in Section 2.6 of~\cite{jakob_integer_hyperflows}.

In Figure~\ref{fig: hypergraph}, three possible hyperflows in a hypergraph are depicted.
The unboxed integers are the flow values on the half-edges, while the boxed integers are the flow values for hyperedges in $E$.
For clarity, only half-edges with non-zero flow values are shown.

One key feature of integer hyperflows is that they give a flexible and generic way to \emph{specify} pathways via constraints on the flow on half-edges.
This may be as straightforward as specifying an overall reaction for the pathway by fixing the flow on the half-edges for the molecules in the source and target sets and setting the flow to zero for the rest of the half-edges.
Or one may specify upper and lower limits for some half-edges and free values for others,
for instance to ask if there is any way to produce at least some target molecules using any amount of source molecules,
while at the same time allowing waste products.

Then a \emph{search} for pathways fulfilling the specification amounts to finding flow vectors~$f$ obeying those constraints, in addition to the constraint in Equation~\eqref{int_hyperflow_jakob}.
Since integer hyperflows are integer valued, it is natural to model the situation with an ILP and search for~$f$ using an ILP solver.
This idea has been implemented as part of the software system~M{\O}D~\cite{mod_web}.
The result is a powerful and flexible tool for specifying and searching for pathways (integer hyperflows) in a given reaction network (hypergraph).
The flexibility is further increased by the possibility of having flow values in the ILP objective function such that flow solutions minimizing side products can be found.
The flow values on hyperedges can enter the specification and/or optimization, allowing for the search for solutions to minimize the use of a given reaction deemed particularly expensive to include in the pathway.

\section{Methods}\label{sec: methods}

We aim to incorporate elements from thermodynamics to refine the existing search query for pathways in reaction networks.
Analogous to typical thermodynamic problems, where pathways maximizing the work done by the flow of heat between the system and the reservoir are sought (such as the cycle of a heat engine),
we seek pathways in our system to maximize the chemical work done (interpreted as maximizing the production of the target molecules) by directing the flow of molecules along specific reactions on the free energy landscape.

Classical thermodynamics---contrary to the `dynamics' in the name---is used to study systems at equilibrium,
or near equilibrium, when the thermodynamic driving forces are in exact balance.
In a closed system, the concentrations of the molecules in the system would be such that the chemical potentials of all the molecules in the reaction network would be the same.
To achieve the required flow along the pathway returned by our method, one has to perturb the system from equilibrium by tweaking the molecular concentrations in the directions returned by the solution.
In practical scenarios, this is done by supplying a steady inflow of the source molecules and removing the target molecules from the system to maintain a non-zero thermodynamic driving force along the pathway.  
For perturbations that are small enough, the equations from classical thermodynamics describing the equilibrium situation can still be used as a first-order approximation.
The gradient of the chemical potential in general, or in our case, the difference in the chemical potentials of the target molecules and the source molecules in an elementary reaction serves as the thermodynamic force driving a reaction.
In the linear regime, the flow for each reaction would be directly proportional to the thermodynamic forces driving the flow.
The forces driving the flows in a reaction network might not remain linear functions of the difference in the chemical potentials farther from the equilibrium, resulting in non-linear dynamics.
Therefore, in this work, we would restrict ourselves to analyzing the system perturbed, within acceptable limits,
from the equilibrium, so that using equations from equilibrium thermodynamics does not introduce large errors into the calculations.

\subsection{Mathematical Formulation}\label{math_form}

The first step to make the search more realistic is to assign the chemical potentials to the vertices and add constraints to allow only thermodynamically favorable reactions in the returned pathway. 
Chemical potentials are real-valued numbers. 
Hence, the inclusion of customized linear constraints and objective functions involving floating-point variables for the chemical potentials and concentrations of molecules requires us to extend the existing ILP model in M{\O}D to one using MILP incorporating these variables. This is the key focus of the current work.

MILP attempts to model the system with linear inequalities involving some variables (both integers and floats, hence mixed) that describe the system. An infeasible system of inequalities indicates that there is no solution to the problem under those specific constraints. When a feasible solution space exists, one often attempts to optimize the solution by maximizing (or minimizing) the value of some objective function. In Section \ref{milp_implication}, we develop the structure of the MILP model using logical constraints using implications, followed by linearizing it so that it can be solved using a solver in Section \ref{milp_linearize}.

\subsubsection{Constraint-Based Model}
\label{milp_implication}

Since we are extending the existing integer hyperflow formulation defined earlier, the base ILP formulation for modeling flow conservation in pathways in reaction networks is inherited from that work.
\begin{equation}
\forall v\in V\colon \sum_{ e\in \overline{E} } s_{ve}^+ f_e - \sum_{ e\in \overline{E} } s_{ve}^- f_e = 0
\tag{\ref{int_hyperflow_jakob}}
\end{equation}
Any set of integer flow values on $E$ induces some overall reaction, which corresponds to flow values on $E^+ \cup E^-$ that make Equation~\eqref{int_hyperflow_jakob} hold.
However, the set of available source molecules is often constrained by practical considerations such as cost, ease of availability, and purity.
Similarly, for the flow to be of practical utility, we could be interested in flows which maximize the outflow of a particular molecule, produce minimum byproducts, and perhaps limit the use of particularly expensive reactions.
Such conditions are modeled as constraints on the flow values $f_e$, in particular for $e$ in $E^+$ and $E^-$, but potentially also in $E$.
These constraints constitute a structural specification of the pathway searched for:
\begin{align}
    \text{Structural specification constraints on $f_e$ for } e \in \overline{E} \label{inflow_outflow_constraints}
\end{align}

To incorporate thermodynamics into the modeling, we assign a chemical potential for each vertex $v$, $x^{G^0}_v$, under standard physical conditions,
which are calculated using a thermodynamic oracle described later.
The differences in the chemical potentials of the target set and the source set in a reaction $e$, $x^{\Delta G}_e$, serve as the thermodynamic force driving the reaction, dictating the direction of flow of matter along that hyperedge.
In the linear regime, not far away from the equilibrium, the thermodynamic flow on a particular hyperedge is directly proportional to the thermodynamic force driving it.
The difference in chemical potential for a reaction can be manipulated by altering the concentrations of the molecules involved. 
We choose to represent the log concentration of a molecule $v$ by the variable $x_v^K$.
The nudged difference in chemical potentials for a reaction $e$ is represented as the variable $x^{\Delta G}_e$, after taking into account the effect of the log concentration $x_v^K$, of the molecules involved in the reaction $e$.
We would like to have an assignment of variables $x_v^K$ (within practical limits) such that the pathway consists of reactions with only negative differences in chemical potentials, in other words, thermodynamically favorable reactions.  

The flow $f_e$ through a hyperedge $e$ must be positive:
\begin{equation}
    \forall e \in \overline{E}\colon  0 \leq f_e
    \label{proposed_constraint2}
\end{equation}
Since the reactions are reversible, separate hyperedges are added for the forward and backward directions.
Under specific physical conditions, thermodynamic forces dictate that either the forward or the backward direction of a reversible reaction is more likely.
The direction of flow is in the direction in which the difference in the chemical potential of a hyperedge $e$, $x^{\Delta G}_e$, is negative.
The following implication constraint ensures that the pathway returned by the MILP solver consists of thermodynamically favorable reactions:
\begin{equation}
    \forall e \in E\colon  f_e > 0 \implies x^{\Delta G}_e \leq 0 \label{proposed_constraint3}
\end{equation}
The difference in chemical potential, $x^{\Delta G}_e$ under arbitrary log concentration values, $x^K_v$ of the molecules participating in the reaction $e$ is calculated from the difference in chemical potential under standard conditions, $x^{G^0}_v$, by the equations:
\begin{align}
    \forall e&\equiv (e^-, e^+) \in E\colon  \nonumber\\
    x^{\Delta G}_e &= \left(\sum_{v\in e^+} x^{G^0}_v - \sum_{v\in e^-} x^{G^0}_v\right) \nonumber\\
    &+ RT\cdot \left(\sum_{v\in e^+} x^K_v - \sum_{v\in e^-} x^K_v\right) \label{proposed_constraint4}
\end{align}
The molecules on the left and right sides of an elementary reaction belong to $e^-$ and $e^+$ respectively.
$R$ is the universal gas constant (in appropriate units) while $T$ is the absolute temperate of the system.
Therefore, $RT$ is a constant for the system.  
The log concentration of the molecules, $x_v^K$, is allowed to vary within a range $\left[x^K_v\right]_{\text{min}} \leq x_v^K \leq  \left[x^K_v\right]_{\text{max}}$,
set by practical and physical constraints detailed in Appendix \ref{second appendix}, to mimic the conditions inside a reactor.
\begin{equation}
    \left[x^K_v\right]_{\text{min}} \leq x^K_v \leq \left[x^K_v\right]_{\text{max}} \label{proposed_constraint8}
\end{equation}

Note that the dependence on the difference in the chemical potentials of a reaction $e$ on the concentrations of the molecules involved is not linear.
\begin{align*}
    x^{\Delta G}_e &= x^{\Delta G^0}_e + RT \ln\left(\frac{\prod_{v\in e^+} \text{conc}(v)}{\prod_{v\in e^-} \text{conc}(v)}\right) \\
    &= x^{\Delta G^0}_e + \frac{RT}{\ln 10}\left(\log\left(\prod_{v\in e^+} \text{conc}(v)\right)\right.\\
    & \text{\hspace{3cm}}- \left.\log\left(\prod_{v\in e^-} \text{conc}(v)\right)\right) \\
    &= x^{\Delta G^0}_e + \frac{RT}{\ln 10}\left(\sum_{v\in e^+} \log \text{conc}(v)\right.\\
    & \text{\hspace{3cm}}- \left.\sum_{v\in e^-} \log \text{conc}(v)\right) \\
    \text{where, }\\
    x^{\Delta G^0}_e &= \sum_{v\in e^+} x^{G^0}_v - \sum_{v\in e^-} x^{G^0}_v
\end{align*}
However, it can be represented as a linear expression, if the variable is chosen to be the log concentration of the corresponding molecule $v$, $x_v^K \equiv \log_{10} \text{conc}(v)$, in the linear program.

\subsubsection{Objective Function}

Formulating the search for pathways in reaction networks as an MILP allows us to add an implication constraint to rule out pathways containing one or more individual reactions not favored by thermodynamics.
However, it enables us to exploit another feature of MILPs,
namely the objective function to quantify the likelihood of the pathway.

We would like to formulate the search for pathways as an optimization problem aimed at maximizing the outflow of the target molecule(s).
To maximize the intended objective, we attempt to maximize the net thermodynamic force driving the flow.
Considering the net thermodynamic forces driving the flow to be the sum of the thermodynamic driving force for each individual reaction in the pathway,
we choose as objective function the minimization of the cumulative sum of the difference in chemical potentials for each constituent reaction with a non-zero flow. 
\begin{equation}
    \max \left(\sum_{e\colon f_e > 0} x^{\Delta G}_e \right)
    \label{objective function}
\end{equation}
This choice of the objective function \eqref{objective function} is a heuristic that fits well with the cost of perturbing the system so that the queried flow is effected with all reactions being thermodynamically driven in the required direction.

The motivation behind choosing this particular objective function can also be argued for from the perspective of maximizing the probability of the pathway.
We interpret the free energy difference to be the measure of the probability of the reaction as
\begin{equation}
    p(e) \sim \exp\left(-x^{\Delta G}_e\right)
    \label{prob_transition}
\end{equation}
Assuming individual reactions in the pathway as independent events,
the probability that the sequence of reactions in the pathway occurs is given by
\begin{align*}
    p\left(\bigwedge_{e\colon f_e > 0} e\right) &= \prod_{e: f_e > 0} p(e) \\
    &\sim \prod_{e\colon f_e > 0} \exp\left(-x^{\Delta G}_e\right) \\
    &= \exp\left(- \sum_{e: f_e > 0} x^{\Delta G}_e\right)
\end{align*}
Since $\forall (x, y)\colon  x\leq y \iff \exp(x) \leq \exp(y)$,
maximizing the probability of the pathway is equivalent to minimizing the objective function (Equation~\eqref{objective function}).

\subsubsection{Temporal Ordering of Molecules}

The ILP formulation of the problem~\cite{jakob_integer_hyperflows} accounts for the conservation of the for each molecule,
but not for temporal ordering of the reactions in the returned flow.
In scenarios such as synthesis planning~\cite{neurips_dag, rolf_synthesis_planning},
we would like to interpret a pathway as a directed acyclic process that generates the target molecule(s).
We therefore introduce constraints to enforce a partial temporal order on the found pathways.

The ordering is implemented by introducing a supplementary integer variable $t_v$ for each vertex $v$ that keeps track of their ordering in the pathway.
The condition $t_{u} < t_{v}$ implies that the vertex $u$ is visited by the flow before the vertex $v$.
This is ensured by the following implication constraint:
\begin{align}
    \forall e\equiv (e^-, e^+)&\in E\colon \nonumber\\
        f_e &\geq 1 \implies t_v \geq t_u + 1, \label{prevent_back_edges}\\
        \forall &\text{ pairs } (u, v) \text{ such that } u \in e^-, v\in e^+ \nonumber
\end{align}
That is, the constraint ensures that reactants are assigned a lower order than products of a reaction with flow. 
We acknowledge that it can be too restrictive to enforce constraints to return pathways with an assigned temporal ordering of vertices because they would eliminate all pathways with cycles in them.
Cycles are present whenever catalysts are used and in numerous metabolic networks.
We merely suggest the possibility of adding such constraints as part of the larger MILP model.
There might be other use cases where these temporal constraints should be disabled in the model.
In these cases, the realizability question for pathways can be addressed using a separate,
more robust tool using the computationally more involved formalism of Petri nets on top of the MILP solver.
We refer the reader to a detailed study~\cite{sissel_realisability_petrinets} for a more thorough treatment of this issue of realizability of pathways.

\subsection{Implementation}
\label{milp_implement}

The objective function \eqref{objective function} is a conditional sum: the value of the variable $x^{\Delta G}_e$ is added to the objective function only if the corresponding hyperedge is used in the pathway.
It needs to be converted into a simple linear sum of a set of MILP variables before the MILP solver can evaluate it.  
In addition, constraints \eqref{proposed_constraint3} and \eqref{prevent_back_edges} in the mathematical formulation of the problem have been specified as implications. 
They need to be linearized before fed into the MILP solver.
This section describes these implementation details for the MILP model formulated in Section \ref{milp_implication}.

\subsubsection{Implementing the Objective function}

The objective function \eqref{objective function} minimizes the cumulative sum of the differences in chemical potentials along the pathway
$\min\left(\sum_{e\colon f_e > 0} x^{\Delta G}_e \right)$,
rather than the simple sum of the differences in chemical potentials.
For all reactions $e$, $\min\left(\sum_{e \in E} x^{\Delta G}_e \right)$, in the reaction network. 
For each hyperedge $e$, a binary variable $z_e \in \{0, 1\}$ is introduced,
indicating whether that hyperedge is used in the pathway, by the following constraints:
\begin{align}
    z_e = 0 &\iff f_e = 0 \label{biimp3} \\
    z_e = 1 &\iff f_e > 0 \label{biimp4}
\end{align}
We define an additional variable $\overline{x}^{\Delta G}_e$ for each hyperedge, which is assigned by the following constraints:
\begin{align}
    \overline{x}^{\Delta G}_e = 0 &\iff z_e = 0 \label{biimp1} \\
    \overline{x}^{\Delta G}_e = x^{\Delta G}_e &\iff z_e = 1 \label{biimp2}
\end{align}
The objective function is defined as the sum of these $|E|$ variables for all hyperedges in the hypergraph;
$\min\left(\sum_{e \in E} \overline{x}^{\Delta G}_e \right)$.
Hyperedges that are not utilized by the pathway have $\overline{x}^{\Delta G}_e = 0$ and would not contribute to the value of the objective function, implementing the objective function as a linear expression.

\subsubsection{Linearizing the Constraints}
\label{milp_linearize}

The bi-implication constraints, Equations~\eqref{biimp3} and \eqref{biimp4}, associating the boolean variables $z_e$ to the edges $e$ are implemented using the big-M method ~\cite{bigM},
where $M > 0$ is a large positive integer constant.
\footnote{The constant $M$ has to be larger than the possible values of the variables in the MILP formulation which include the flows $f_e$ for the hyperedges and the chemical potentials $x^{G^0}_v$ for the vertices (in atomic units). In our current implementation $M = 100$ suffices.}
\begin{align}
    \forall e \in E\colon & \nonumber \\
    z_e &\leq f_e \label{linear_constraint6}\\
    M\cdot z_e &\geq f_e \label{linear_constraint7}
\end{align}
The expansion of how the added linear constraints in Equations~\eqref{linear_constraint6} and \eqref{linear_constraint7}
lead to the assignment of $z_e$ that satisfies the implication constraints in Equations~\eqref{biimp3} and \eqref{biimp4}
is shown in Table~\ref{tab: truth_table_constraint1}.
\begin{table}
    \centering
    \begin{tabular}{@{}c@{\quad}c@{\qquad}c@{\quad}c@{}}
        \toprule
        Case of $f_e$ & \eqref{linear_constraint6} & \eqref{linear_constraint7} & Implication for $z_e$ \\
        \midrule
        $f_e = 0$ & $z_e \leq 0$ & $M\cdot z_e \geq 0$ & $z_e = 0$ \\
        $f_e = k > 0$ & $z_e \leq k$ & $M\cdot z_e \geq k$ & $z_e = 1$ \\
        \bottomrule
    \end{tabular}
    \caption{The linear constraints in Equations \eqref{linear_constraint6} and \eqref{linear_constraint7}
    are equivalent to the bi-implication constraints in Equations \eqref{biimp3} and \eqref{biimp4}.}
    \label{tab: truth_table_constraint1}
\end{table}

The implication constraint in Equation~\eqref{proposed_constraint3} determines whether a hyperedge can be used in the flow depending on the differences in the chemical potentials for the hyperedges $x^{\Delta G}_e$.
This is linearized as follows:
\begin{align}
    \forall e \in E\colon & \nonumber \\
    x^{\Delta G}_e + M\cdot (z_e - 1) &\leq 0 \label{linear_constraint1}
\end{align}
The equivalence of the added linear constraints in Equation~\eqref{linear_constraint1} with the implication constraint in Equation~\eqref{proposed_constraint3} is shown in Table~\ref{tab: truth_table_constraint2}.
\begin{table*}[ht]
    \centering
    \begin{tabular}{@{}c@{\qquad}c@{\qquad}c@{\qquad}c@{\qquad}c@{}}
        \toprule
        Case of $f_{e}$ & Implication for $z_{e}$ & Case of $x^{\Delta G}_e$ & \eqref{proposed_constraint3} satisfied? & \eqref{linear_constraint1} satisfied? \\
        \midrule
        $f_{e} = 0$ & $z_{e} = 0$ & $x^{\Delta G}_e > 0$ & True & True \\
        $f_{e} = 0$ & $z_{e} = 0$ & $x^{\Delta G}_e \leq 0$ & True & True \\
        $f_{e} > 0$ & $z_{e} = 1$ & $x^{\Delta G}_e > 0$ & False & False \\
        $f_{e} > 0$ & $z_{e} = 1$ & $x^{\Delta G}_e \leq 0$ & True & True \\
        \bottomrule
    \end{tabular}
    \caption{The linear constraint in Equation~\eqref{linear_constraint1} is equivalent to the implication constraint in Equation~\eqref{proposed_constraint3}.}
    \label{tab: truth_table_constraint2}
\end{table*}

The bi-implication constraints in Equations~\eqref{biimp1} and \eqref{biimp2} assign the variable $\overline{x}^{\Delta G}_e$ with the differences in the chemical potentials for the hyperedges for the evaluation of the objective function.
Using the variables $z_e$, we implement the constraints as
\begin{align}
    \forall e \in E\colon & \nonumber \\
    x^{\Delta G}_e - \overline{x}^{\Delta G}_e &\leq M(1 - z_e) \label{linear_constraint2} \\
    x^{\Delta G}_e - \overline{x}^{\Delta G}_e &\geq -M(1 - z_e) \label{linear_constraint3} \\
    \overline{x}^{\Delta G}_e &\leq M\cdot z_e \label{linear_constraint4} \\
    \overline{x}^{\Delta G}_e &\geq -M\cdot z_e \label{linear_constraint5}
\end{align}
The expansion of how the added linear constraints in Equations~\eqref{linear_constraint2}, \eqref{linear_constraint3}, \eqref{linear_constraint4}, and \eqref{linear_constraint5} lead to the desired assignment of $\overline{x}^{\Delta G}_e$ satisfying the implication constraints in Equations~\eqref{biimp1} and \eqref{biimp2} is shown in Table~\ref{tab: truth_table_constraint3}.
\begin{table*}[ht]
    \centering
    \begin{tabular}{@{}c@{\qquad}c*{4}{@{\qquad}c}l@{\qquad}c@{}}
        \toprule
        &&\multicolumn{4}{c}{Constraint on $\overline{x}^{\Delta G}_e$ by}\\
        \cmidrule{3-6}
        Case of $z_{e}$ & Case of $x^{\Delta G}_e$ &
            \eqref{linear_constraint2} & \eqref{linear_constraint3} & \eqref{linear_constraint4} & \eqref{linear_constraint5} &
        & Implication for $\overline{x}^{\Delta G}_e$ \\
        \midrule
        $z_{e} = 0$ & $x^{\Delta G}_e > 0$ & & & $\leq 0$ & $\geq 0$ && $=0$ \\
        $z_{e} = 0$ & $x^{\Delta G}_e \leq 0$ & & & $\leq 0$ & $\geq 0$ && $=0$ \\
        $z_{e} = 1$ & $x^{\Delta G}_e > 0$ &
            \multicolumn{6}{c}{
                \raisebox{0.5ex}{\rule{8.5em}{0.4pt}}
                \enspace Cannot happen \enspace
                \raisebox{0.5ex}{\rule{8.5em}{0.4pt}}
                } \\
        $z_{e} = 1$ & $x^{\Delta G}_e \leq 0$ & $\geq x^{\Delta G}_e$ & $\leq x^{\Delta G}_e$ & & && $= x^{\Delta G}_e$ \\
        \bottomrule
    \end{tabular}
    \caption{The linear constraints of Equation~\eqref{linear_constraint2}, \eqref{linear_constraint3}, \eqref{linear_constraint4}, and \eqref{linear_constraint5}
    assign $\overline{x}^{\Delta G}_e$ according to the bi-implication constraints in Equations~\eqref{biimp1} and \eqref{biimp2}.
    Note that the third case cannot happen due to Equation~\eqref{proposed_constraint3},
    because $x^{\Delta G}_e > 0 \implies f_e = 0 \implies z_e = 0$.}
    \label{tab: truth_table_constraint3}
\end{table*}

The implication constraint in Equation~\eqref{prevent_back_edges} assigns the temporal ordering $t_v$ to each vertex $v$.
It is implemented as a linear constraint as
\begin{align}
    \forall \text{ pairs } (u, v)\,\,\,|\,\,\, \exists e\equiv (e^-, e^+) \text{ with } u &\in e^-, v \in e^+\colon \nonumber \\
    t_v - t_u - |V| (z_e - 1) - 1 &\geq 0 \label{linear_constraint8} \\
    \forall v\in V,\text{\hspace{1cm}}& \nonumber \\
    0 \leq t_v &\leq |V| \label{linear_constraint9}    
\end{align}
The constraint in Equation~\eqref{linear_constraint8} enforces the assignment $t_{v} > t_{u}$ only if the corresponding hyperedge is used in the hyperflow.
The size of the set of vertices in the hypergraph $|V|$ is used instead of the large integer $M$ in the constraint in Equation~\eqref{linear_constraint8} and as an upper bound for the temporal ordering variable $t_v$ for each vertex $v$ in the constraint in Equation~\eqref{linear_constraint9}.
The number of constraints added to the MILP model to determine the topological sorting of the vertices is $\mathcal{O}\left(|V|^2\right)$.

\subsection{Implemented Model}\label{implemented model}

For an overview, the entire set of variables in the model is listed in Table~\ref{tab: MILP variables} and constants in Table~\ref{tab: MILP constants}. The mixed integer linear program developed in the earlier sections and given to the MILP solver is summarized here.

\begin{align}
    & \min\left(\sum_{e \in E} \overline{x}^{\Delta G}_e \right) \label{obj_func_impl} \\    
    \forall v\in V\colon  & \sum_{ e\in \overline{E} } s_{ve}^+ f_e - \sum_{ e\in \overline{E} } s_{ve}^- f_e = 0 \tag{\ref{int_hyperflow_jakob}} \\
    \text{Structural speci}&\text{fication constraints on $f_e$ for } e \in \overline{E} \tag{\ref{inflow_outflow_constraints}}
\end{align}
\begin{align}
    \forall e&\equiv (e^-, e^+) \in E\colon \nonumber\\
    0 &\leq f_e  \tag{\ref{proposed_constraint2}} \\
    0 &\geq x^{\Delta G}_e + M\cdot (z_e - 1) \tag{\ref{linear_constraint1}} \\
    x^{\Delta G}_e &= \left(\sum_{v\in e^+} x^{G^0}_v - \sum_{v\in e^-} x^{G^0}_v\right) \nonumber\\
    & \hspace{0.5cm}+ RT\cdot \left(\sum_{v\in e^+} x^K_v - \sum_{v\in e^-} x^K_v\right) \tag{\ref{proposed_constraint4}} \\
    M(1 - z_e) &\geq x^{\Delta G}_e - \overline{x}^{\Delta G}_e  \tag{\ref{linear_constraint2}} \\
    -M(1 - z_e) &\leq x^{\Delta G}_e - \overline{x}^{\Delta G}_e \tag{\ref{linear_constraint3}} \\
    M\cdot z_e &\geq \overline{x}^{\Delta G}_e \tag{\ref{linear_constraint4}} \\
    -M\cdot z_e &\leq \overline{x}^{\Delta G}_e \tag{\ref{linear_constraint5}} \\
    \forall (u, v)&| u \in e^-, v \in e^+\colon \nonumber\\
    0 &\leq t_v - t_u - |V| (z_e - 1) - 1 \tag{\ref{linear_constraint6}} \\
    \forall v \in V\colon  & \nonumber\\
    0 &\leq t_v \leq |V| \tag{\ref{linear_constraint7}} \\
    \left[x^K_v\right]_{\text{min}} &\leq x^K_v \leq \left[x^K_v\right]_{\text{max}} \tag{\ref{proposed_constraint8}}
\end{align}

\begin{table*}[ht]
    \centering
    \begin{tabular}{@{}c@{\qquad}l@{\qquad}ll@{}}
        \toprule
        Constant & datatype & & description \\
        \midrule
        $M$ & integer & & a big integer used for linearizing the constraints \\
        & & & (type promoted to float, if required) \\
        $x^{G^0}_v$ & float & & chemical potential of a vertex $v$ under standard physical \\
        & & & conditions as assigned by the thermodynamic oracle \\
        $x^{\Delta G^0}_e$ & float & & differences in chemical potential of in-vertices and out-vertices \\
        & & & of a hyperedge $e$ under standard physical conditions \\
        \bottomrule
    \end{tabular}
    \caption{Constants in the proposed MILP formulation.
    The constants $M$ and $x^{G^0}_v$ are provided as input,
    while the constants $x^{\Delta G^0}_e$ are calculated based on the given $x^{G^0}_v$.
    }
    \label{tab: MILP constants}
\end{table*}%
\begin{table*}[ht]
    \centering
    \begin{tabular}{@{}c@{\qquad}l@{\qquad}ll@{}}
        \toprule
        MILP variable & datatype & & description \\
        \midrule
        $f_e$ & integer & & flow variable for a hyperedge $e$, non-negative \\
        $z_e$ & boolean & & indicator variable for a hyperedge $e$ \\
        & & & denoting if it is used in the flow \\
        $x^{\Delta G}_e$ & float & & differences in chemical potential of in-vertices \\
        & & & and out-vertices of a hyperedge $e$ \\
        & & & after accounting for their concentrations \\ 
        $\overline{x}^{\Delta G}_e$ & float & & copy of $x^{\Delta G}_e$ used in the (linearized) objective function \\
        $x_v^K$ & float & & log concentration of a vertex $v$\\
        $t_v$ & integer & & topological ordering number for vertex $v$, non-negative\\
        \bottomrule
    \end{tabular}
    \caption{Variables in the proposed MILP formulation.}
    \label{tab: MILP variables}
\end{table*}%

The computational complexity of finding MILP solutions is in general NP-hard~\cite{Karp:72},
and so is the restricted case of finding chemical pathways~\cite{integer_hyperflow_np_hard}.
Despite this worst-case complexity, modern MILP solvers efficiently handle most practically relevant instances in reasonable time.
From a practical perspective, it is often valuable to explore \emph{multiple} optimal or near-optimal solutions---however, MILP solvers generally provide the optimal solution.
Some solvers, e.g., IBM ILOG CPLEX, do provide a so-called solution pool facility for enumerating multiple solutions,
but there is little user control over how it is populated, in particular how solutions are compared for equality.

The \texttt{thermodynamics module} described in this contribution is implemented as an extension of the larger software package M{\O}D~\cite{jakob_software}, available at \url{http://mod.imada.sdu.dk/}.
The package offers the option of using multiple MILP solvers, currently IBM ILOG CPLEX, Gurobi, and Coin-OR CBC.
For the results presented, we have used Coin-OR CBC~\cite{cbc,cbc2}.
M{\O}D implements a MILP enumeration algorithm on top of the solver in use.
A set of integer or binary variables to enumerate over can be specified.
This enumeration in M{\O}D proceeds as tree search algorithm on the domains of this set of variables,
with calls to the MILP solver evaluating the feasibility of obtaining optimal solutions to the subproblems.

In the MILP model presented, there are variables with a continuous domain to represent the concentrations and chemical potentials of the molecules.
We do not consider minor changes in these variables to represent a new solution, as long as the underlying flow is unchanged.
Hence, the enumeration is over the variables $f_e,\forall e\in E$.
The resulting output is a list of different integral flows,
listed in the order of increasing value of the objective function \eqref{obj_func_impl}.

\subsection{Thermodynamic Oracle}
\label{thermodynamic_oracle}

Given a reaction network, we assign chemical potentials to the molecules using what we had earlier referred to as a `thermodynamic oracle'. Using quantum mechanical methods based on DFT as this `thermodynamic oracle' quickly becomes impractical as the number of electrons in the molecule (denoted by $n$) increases.
Calculating the electron density in a molecule using DFT is an iterative process. Each step invokes gradient descent, which scales beyond $\mathcal{O}(n^3)$ for classical DFT based methods for a single-point calculation because it involves matrix inversions. After the electron density for the molecule is optimized, it can be used to calculate the electronic contribution to the energy of a molecule.
Machine-learned relations for potential assignment exist; ~\cite{NN_molecule_energies} require optimized atomic coordinates to determine potentials, while ~\cite{ml_energy,ml_energy1,ml_energy2,ml_energy3} utilize the connectivity of the molecular graph or even a string representation of it~\cite{ml-energy4}. However, these machine-learned relations might not be well generalizable; they could perform poorly when molecules sufficiently different from those in the dataset are encountered~\cite{ml_energy5,ml_energy6,ml_energy7}. Ideally, kinetic rate parameters would have to be incorporated into the search for pathways in a reaction network as integer flows because they dictate which reactions actually occur in the system and how fast. However, calculating activation free energies using DFT-based quantum mechanical methods to determine these kinetic parameters would demand more time and computational resources.

As an alternative, this work adopts a compromise oracle based on equilibrium thermodynamics. This approximation has some theoretical basis as the Evans–Polanyi–Semenov principle states that within a class of reactions, the thermodynamics of the reaction acts as a proxy for the kinetics of the reaction. It allows us to substitute the energy barrier for a reaction with its free energy difference, resulting in the relation in Equation~\eqref{prob_transition}. Specifically, to balance accuracy, efficiency, generalizability, and running time, we employ a semi-empirical method, GFN2-xTB~\cite{grimme_xtb}, for potential assignment. Although this method compromises accuracy compared to DFT, the errors are acceptable to demonstrate the utility of the mathematical model on an example reaction network.
We detail the thermodynamic oracle to assign chemical potentials to the molecules $x_v^{G^0}$, in the reaction network in Appendix \ref{first appendix}.
An advantage of this modular approach is that it provides the user freedom to substitute the proposed thermodynamic oracle with another tool, for example, pretrained neural networks, mapping a molecule to its chemical potential (or even the reactions to their kinetic parameters) depending on the desired accuracies and available computational resources.

\section{Results}\label{sec: results}

In this section, we demonstrate the method on a reaction network involving hydrogen cyanide, formamidic acid, and formamide, which has been studied in~\cite{implemented_CRN}.
The reaction network was generated using M{\O}D by the expansion of the molecular space with the input molecules \chemfig{HCN}, \chemfig{NH_3}, and \chemfig{H_2O}, by the recursive application of the graph transformation rules listed in Appendix \ref{third appendix}.
In the graphs in Appendix \ref{third appendix}, individual molecules are represented as undirected labeled graphs (with atoms represented as vertices, labeled by the element symbol, and bonds are represented as edges, labeled by bond types).
Stereoinformation is not represented.
Molecular graphs also constitute the vertices of the hypergraph representing the reaction network. The graph transformation rules in Appendix \ref{third appendix} represent classes of reactions, i.e., reaction templates. The left-hand side of the rules is a subgraph, which may be substituted by the right-hand side subgraph, when a match has been found in a molecular graph present in the reaction network. This executes a reaction, thereby expanding the molecular space. Following~\cite{implemented_CRN}, the expansion was restricted to only generate molecules with a maximum of three carbon, three nitrogen, and three oxygen atoms. Moreover, only stable molecules ($x_v^{G^0} \leq 0$) were kept. 
The generated reaction network consists of $67$ vertices and $202$ hyperedges, and is shown in Figure~\ref{fig: hypergraph_for_RN} for completeness.
A list of all molecules and reactions in the reaction network can be found in the associated GitHub repository~\cite{data_repository}.
An estimate of the atomic coordinates of the molecules was deduced from their connectivity using Open Babel~\cite{open_babel}.
Using xTB~\cite{grimme_xtb}, this geometry was then refined and used to calculate the chemical potentials under standard conditions with water as the solvent.

\begin{figure}[t!]
    \centering
    \begin{subfigure}{0.5\textwidth}
        \centering
        \includegraphics[width=\textwidth]{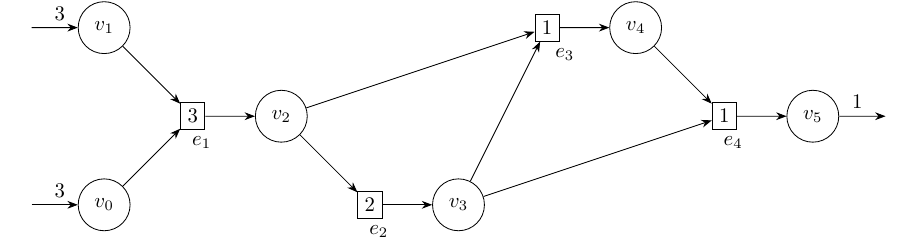}
        \caption{The conventional pathway from previous studies of the reaction network from \cite{implemented_CRN}.\label{pathway1}}
    \end{subfigure}
    
    \begin{subfigure}{0.5\textwidth}
        \centering
        \begin{scriptsize}
        \begin{tabular}{@{}ccccr@{}}
            \\
            \toprule
            vertex &  & molecular structure & & in \cite{implemented_CRN} \\
            \midrule
            $v_0$ && \tiny{HCN} && $1$ \\
            $v_1$ && \tiny{H$_2$O} && not numbered \\[0.1cm]
            $v_2$ && \tiny{\chemfig{HO-[:30]=_[:-30]NH}} && $4$ \\[0.15cm]
            $v_3$ && \tiny{\chemfig{O=_[:30]-[:-30]NH_2}} && $5$ \\[0.2cm]
            $v_4$ && \tiny{\chemfig{O=_[:30]-[:-30]N(-[::-60,0.4,,,draw=none]H)-[:30](-[:90]OH)-[:-30]NH_2}}&& $16$ \\[0.25cm] 
            $v_5$ && \tiny{\chemfig{O=_[:30]-[:-30]N(-[::-60,0.4,,,draw=none]H)-[:30](-[:90]OH)-[:-30]N(-[::-60,0.4,,,draw=none]H)-[:30](-[:90]OH)-[:-30]NH_2}} && $17$ \\[0.2cm] 
            \bottomrule
        \end{tabular}
        \end{scriptsize}
        \caption{List of molecules used in the pathway.\label{tableOfMolecules1}}
    \end{subfigure}
    
    \begin{subfigure}{0.5\textwidth}
        \centering
        \begin{scriptsize}
        \begin{tabular}{@{}c@{\phantom{aa}}c@{\phantom{aa}}c@{\phantom{aa}}rrc@{}}
            \\
            \toprule
            vertex & inFlow & outFlow & $x_v^{\Delta G}$ & $x_v^K$ & $t_v$ \\
            \midrule
            $v_0$ & $3$ & $0$ & $-5.507$ &  $1.00$ & $0$ \\
            $v_1$ & $3$ & $0$ & $-5.068$ & $1.00$ & $0$ \\
            $v_2$ & $0$ & $0$ & $-10.610$ & $1.00$ & $1$ \\
            $v_3$ & $0$ & $0$ & $-10.625$ & $1.00$ & $2$ \\
            $v_4$ & $0$ & $0$ & $-21.249$ & $-3.00$ & $5$ \\   
            $v_5$ & $0$ & $1$ & $-31.869$ & $-3.00$ & $8$ \\
            \bottomrule
        \end{tabular}
        \end{scriptsize}
        \caption{Values assigned by the MILP solver the the variables for the vertices.\label{molecules1}}
    \end{subfigure}
    
    \begin{subfigure}{0.5\textwidth}
        \centering
        \begin{scriptsize}
        \begin{tabular}{@{}ccccr@{}}
            \\
            \toprule
            hyperedge & $f_e$ & in($v$) & out($v$) & $\overline{x}^{\Delta G}_e$ \\
            \midrule
            $e_1$ & $3$ & $v_0, v_1$ & $v_2$ & $-94.141$ \\
            $e_2$ & $2$ & $v_2$ & $v_3$ & $-39.628$ \\
            $e_3$ & $1$ & $v_2, v_3$ & $v_4$ & $-8.524$ \\
            $e_4$ & $1$ & $v_3, v_4$ & $v_5$ & $-29.632$ \\
            \midrule
            & & $\sum_e \overline{x}^{\Delta G}_e$ & $=$ & $-171.926$ \\
            & & $\sum_e f_e\cdot \overline{x}^{\Delta G}_e$ & $=$ & $-399.835$ \\
            \bottomrule
        \end{tabular}
        \end{scriptsize}
        \caption{Values assigned by the MILP solver the the variables for the hyperedges.\label{reactions1}}
    \end{subfigure}    
    \caption{\ref{pathway1}: The pathway with the sum of free energies $-171.926$ kJ mol$^{-1}$ and Table \ref{tableOfMolecules1}: the molecules used with the corresponding number in the earlier work~\cite{implemented_CRN}. Table \ref{molecules1} and Table \ref{reactions1}: The MILP solution with the $x_v^{\Delta G}$ values are in atomic units (from xTB) while the $\overline{x}^{\Delta G}_e$ are in kJ mol$^{-1}$. The $x_v^K$ values are the dimensionless equivalent of $\log_{10} \frac{[\text{concentration}]}{1 \text{  mol lit}^{-1}}$.
    }
    \label{fig: conventional_path}
\end{figure}

The primary objective was to study the energetics of the condensation pathways of the smaller input molecules into larger oligomers, with implications for the prebiotic synthesis of amino acids, and to compare it with the pathways reported in~\cite{implemented_CRN}. 
Therefore, the MILP was set up to query the pathways with \chemfig{HCN}, \chemfig{NH_3}, and \chemfig{H_2O} as the source molecules and the trimer of formamide (numbered $17$ in~\cite{implemented_CRN}) as the target molecule.
The MILP formulated to search for pathways with the given input and output vertices had $336$ variables.
As a sanity check, the enumerated list of flow solutions returned by the solver included the pathway suggested in~\cite{implemented_CRN}, although not as the optimal solution when scored according to Equation~\eqref{obj_func_impl}.
This is depicted in Figure~\ref{fig: conventional_path}.
The sum of free energies of reactions in this pathway is $-171.926$ kJ mol$^{-1}$, while the sum of free energies weighted by the number of times it is used in the pathway yields $-399.835$ kJ mol$^{-1}$.
This pathway uses four unique reactions and six different molecules.
Two of them are used multiple times, making seven reactions in total (sum of flows assigned to the hyperedges).

\begin{figure}[t!]
    \centering
    \begin{subfigure}{0.5\textwidth}
        \centering
        \includegraphics[width=\textwidth]{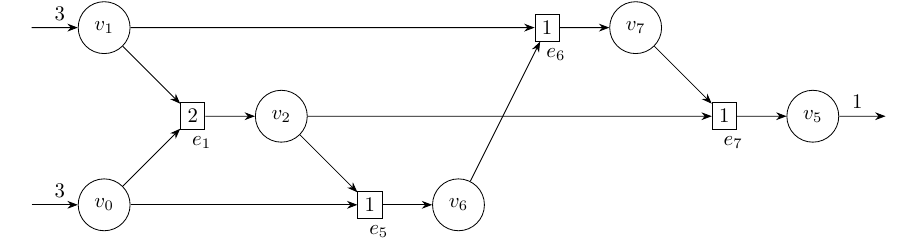}
        \caption{The best-scoring pathway in the considered reaction network.\label{pathway2}}
    \end{subfigure}
    
    \begin{subfigure}{0.5\textwidth}
        \centering
        \begin{scriptsize}
        \begin{tabular}{@{}ccccr@{}}
            \\
            \toprule
            vertex & & molecular structure & & in \cite{implemented_CRN} \\
            \midrule
            $v_0$ && \tiny{HCN} && $1$ \\
            $v_1$ && \tiny{H$_2$O} && not numbered \\[0.1cm]
            $v_2$ && \tiny{\chemfig{HO-[:30]=_[:-30]NH}} && $4$ \\[0.2cm] 
            $v_6$ && \tiny{\chemfig{HO-[:30]=_[:-30]N-[:30]=_[:-30]NH}} && not present \\[0.2cm] 
            $v_7$ && \tiny{\chemfig{HO-[:30]=_[:-30]N-[:30](-[:90]OH)-[:-30]NH_2}} && tautomer of $16$ \\[0.2cm] 
            $v_5$ && \tiny{\chemfig{O=_[:30]-[:-30]N(-[::-60,0.4,,,draw=none]H)-[:30](-[:90]OH)-[:-30]N(-[::-60,0.4,,,draw=none]H)-[:30](-[:90]OH)-[:-30]NH_2}} && $17$ \\[0.2cm] 
            \bottomrule
        \end{tabular}
        \end{scriptsize}
        \caption{List of molecules used in the pathway.\label{tableOfMolecules2}}
    \end{subfigure}
    
    \begin{subfigure}{0.5\textwidth}
        \centering
        \begin{scriptsize}
        \begin{tabular}{@{}c@{\phantom{aa}}c@{\phantom{aa}}crrc@{}}
            \\
            \toprule
            vertex & inFlow & outFlow & $x_v^{\Delta G}$ & $x_v^K$ & $t_v$ \\
            \midrule
            $v_0$ & $3$ & $0$ & $-5.507$ &  $1.00$ & $0$ \\
            $v_1$ & $3$ & $0$ & $-5.068$ & $1.00$ & $0$ \\
            $v_2$ & $0$ & $0$ & $-10.610$ & $1.00$ & $1$ \\
            $v_6$ & $0$ & $0$ & $-16.159$ & $-3.00$ & $2$ \\
            $v_7$ & $0$ & $0$ & $-21.241$ & $1.00$ & $3$ \\        
            $v_5$ & $0$ & $1$ & $-31.869$ & $-3.00$ & $5$ \\
            \bottomrule
        \end{tabular}
        \end{scriptsize}
        \caption{Values assigned by the MILP solver the the variables for the vertices.\label{molecules2}}
    \end{subfigure}
    
    \begin{subfigure}{0.5\textwidth}
        \centering
        \begin{scriptsize}
        \begin{tabular}{@{}ccccr@{}}
            \\
            \toprule
            hyperedge & $f_e$ & in($v$) & out($v$) & $\overline{x}^{\Delta G}_e$ \\
            \midrule
            $e_1$ & $2$ & $v_0, v_1$ & $v_2$ & $-94.141$ \\
            $e_5$ & $1$ & $v_0, v_2$ & $v_6$ & $-122.072$ \\
            $e_6$ & $1$ & $v_1, v_6$ & $v_7$ & $-28.321$ \\
            $e_7$ & $1$ & $v_2, v_7$ & $v_5$ & $-61.162$ \\
            \midrule
            & & $\sum_e \overline{x}^{\Delta G}_e$ & $=$ & $-305.697$ \\
            & & $\sum_e f_e\cdot \overline{x}^{\Delta G}_e$ & $=$ & $-399.837$ \\
            \bottomrule
        \end{tabular}
        \end{scriptsize}
        \caption{Values assigned by the MILP solver the the variables for the hyperedges.\label{reactions2}}
    \end{subfigure}
    
    \caption{\ref{pathway2}: The pathway with minimum sum of free energies ($-305.697$ kJ mol$^{-1}$) returned as the optimal by our model and Table \ref{tableOfMolecules2}: the molecules used with the corresponding number in the earlier work. Table \ref{molecules2} and Table \ref{reactions2}: MILP solution and the assignment of the MILP variables for the optimal pathway with the units same as in Figure \ref{fig: conventional_path}.
    }
    \label{fig: shortest_path}
\end{figure}

Although thermodynamics might provide insight into the equilibrium flows in the reaction network,
it does not describe the path to that equilibrium.
Each reaction has an associated barrier energy that must be overcome for the reaction to occur.
Enumerating pathways using our implementation of the MILP formulation of the problem provides the pathway in Figure~\ref{fig: shortest_path} as one with the minimum value of the objective function ($-305.697$ kJ mol$^{-1}$).
This pathway utilizes six different molecules, four unique reactions, and five reactions in all to realize the pathway, fewer than the previously reported pathway.
Since the difference in the chemical potential between the source and target molecules is fixed,
choosing hyperedges with larger free energy differences would increase the value of the objective function.
Using reactions with larger free energy differences to `cover' the net difference in the chemical potential,
lead to fewer reactions being used in the pathway. 

The main source of difference in the objective function value in the two pathways is the chemical potential of the molecule $v_6$ that is not present in the conventional pathway.
The conventional pathway in~\cite{implemented_CRN} is more symmetric and appealing due to the stepwise nucleophilic attack on the carbonyl \chemfig{C} atom by the amine \chemfig{N} atom of an adjacent $v_3$ molecule that is repeated twice to generate $v_5$.

We observe that the log concentration variables are always assigned extremal values by the constraint in Equation~\eqref{proposed_constraint8}.
This is likely an artifact of using an MILP solver with an objective function for the problem.
An MILP is a convex optimization problem in which minimizing the objective function (involving a subset of the MILP variables) results in a solution located at a vertex of the convex space defined by the constraints.
The bounds for the variables are hand-crafted and it might be more realistic to assign different bounds for the log concentration variable for each vertex.
Alternatively, a better interpretation of the values assigned to the variables $x_v^K$ may be the direction in which the equilibrium concentrations are perturbed to achieve the desired flow.
For example, the concentration values in Table~\ref{molecules2} can be interpreted as the concentrations of $v_6$ and $v_5$ are reduced and those of $v_0$, $v_1$, $v_2$, and $v_7$ are increased to achieve a greater difference in the potentials for a net outflow of the molecule $v_5$ given input molecules $v_0$ and $v_1$. 

\begin{table}[!ht]
    \centering
    \begin{small}
        \begin{tabular}{ccccrr}
            \toprule
            rank & $\sum_{e} f_e$ & $\sum_{e} z_e$ & $|v|$ & \hspace{0.1cm}obj \eqref{obj_func_impl} & $\sum_{e} f_e \cdot x_e^{\Delta G}$ \\
            \midrule
            0 & $5$ & $4$ & $6$ & $-305.697$ & $-399.837$ \\
            1 & $6$ & $5$ & $7$ & $-306.697$ & $-399.837$ \\
            2 & $6$ & $5$ & $7$ & $-305.696$ & $-399.836$ \\

            3 & $6$ & $4$ & $6$ & $-211.556$ & $-399.836$ \\
            4 & $6$ & $4$ & $6$ & $-211.556$ & $-399.836$ \\
            5 & $7$ & $5$ & $7$ & $-211.555$ & $-399.836$ \\        
            6 & $6$ & $4$ & $6$ & $-211.554$ & $-399.836$ \\
            7 & $5$ & $3$ & $5$ & $-211.554$ & $-399.836$ \\
            8 & $7$ & $5$ & $7$ & $-211.554$ & $-399.836$ \\
            9 & $5$ & $3$ & $5$ & $-211.554$ & $-399.836$ \\
            10 & $6$ & $4$ & $6$ & $-211.554$ & $-399.836$ \\
            
            11 & $7$ & $4$ & $6$ & $-171.927$ & $-399.836$ \\
            12 & $7$ & $4$ & $6$ & $-171.926$ & $-399.835$ \\
            \bottomrule
        \end{tabular}
    \end{small}
    \caption{All enumerated pathways to the target molecule in the network.}
    \label{tab: thermo_flow}
\end{table}

An exhaustive search of the reaction space led to the identification of $13$ distinct reaction pathways~\cite{solutions_data}, listed in Table~\ref{tab: thermo_flow} in the order of increasing objective function \ref{obj_func_impl}. The respective columns list the rank, the total number of reactions used, the number of unique reactions used, the number of molecules involved, the objective function \eqref{obj_func_impl} and the sum of the free energy differences of the reactions weighted by the flows. Note that the last column gives the net free energy difference, which is path-independent and therefore the same for all the listed pathways. These pathways were enumerated in $866$ seconds on an AMD$^{\circledR}$ Ryzen 5 PRO 4650U CPU (2.1 GHz) utilizing $8$ threads. 

\section{Discussion}\label{sec: discussions}

In this section, we give a detailed comparison of our methods with some existing frameworks.
Thermodynamics-based metabolic flux analysis (TMFA)~\cite{hatz_thermo_flow} uses linear constraints, but despite similarities in parts of the thermodynamic modeling, the problem addressed is different, with some of the key differences being:
\begin{itemize}
    \item TMFA generates \emph{flux distributions} without thermodynamically infeasible reactions, while our method searches for \emph{explicitly specified pathways} without thermodynamically unfavorable reactions. The flux vector generated by TMFA is real-valued, as opposed to integral flows in the pathways determined by our method. A pathway involves an integer number of times that a reaction is used, which aligns well with integer flows.
    \item TMFA uses a curated dataset of metabolites with chemical potentials determined by group contribution methods. In our case a semi-empirical method (GFN2-xTB) is used to assign chemical potentials to the molecules in the reaction network. However, it can be substituted by any suitable `oracle' that maps a molecule to its Gibbs free energy of the user's choice. This provides flexibility and a wider applicability to this method.
    \item  In this case, a ranking of the enumerated results is possible on the basis of thermodynamics-inspired metrics. Restricting the flow variables to integers allows us to enumerate solutions over the integral variables (i.e., generating not just a single solution but many).
    Additionally, a partial ordering of the vertices can be enforced in the returned integer hyperflow solutions, in order to guarantee their realizability as synthesis plans.
\end{itemize}

We now turn to a comparison with some of the other existing frameworks.
Formulating the problem as an MILP allows us to leverage the many algorithmic advances made in the field of MILP solvers.
This may be one reason why we are able to analyze comparatively larger reaction networks than~\cite{miangolarra_non_ideal_CRN, seifert_thermodynamics_CRN_JCP, nicolis_thermodynamics_CRN,pathway_optimization,pathway_optimization2}.
A further difference from these works is that they utilize kinetic rate equations derived from the law of mass action (reactions under kinetic control), whereas our focus is on the differences in chemical potentials of the molecules in the reaction network near equilibrium (reactions under thermodynamics control).
However, stochastic algorithms offer a computational alternative to analyze even larger reaction networks and query pathways to specific molecules, especially when reaction templates are not known for the classes of reactions of interest~\cite{large_crn,large_crn2,largeCRN_code}.

Weighted directed graphs have been utilized to model the thermodynamic phase space of a reaction network in~\cite{synthesis_planning_CRN_thermodynamics}.
\cite{synthesis_planning_CRN_thermodynamics} uses a constant positive edge weight derived from the Gibbs free energy of the reaction, while we utilize the free energies themselves for a more direct analysis based on thermodynamics.
Several cheminformatics tools and automated protocols to explore reaction networks computationally utilize reactive molecular dynamics simulations~\cite{CRN_MD_tool1} or chemical heuristics~\cite{CRN_kin_tool1, CRN_kin_tool2, CRN_kin_tool3, pickaxe_CRN_analyzer}.
To the best of our knowledge, there does not exist any computational framework for explicitly specifying and searching for pathways in reaction networks while simultaneously incorporating thermodynamic constraints in the present form.

\section{Conclusion} \label{sec: conclusion}

Thermodynamics emerged from efforts to analyze the operation of steam engines while maximizing their work.
In a nice analogy, the search for pathways may be phrased as viewing reaction networks as chemical engines in which we want to maximize the output of some target molecule(s).
Recognizing that thermodynamics, encapsulated by the Gibbs free energies of the reactions, play a pivotal role in determining the probability of a reaction, this work focuses on developing ways to integrate them into the search for pathways.

Exploring pathways with favorable energy profiles involves challenging navigation of a large and complex energy landscape where the driving thermodynamic forces are influenced by the concentration ratios, inflows of the source molecules, and outflows of the target molecules.
This work attempts to address part of that challenge.
However, there might be several refinements, some of which we list here:
\begin{enumerate}
    \item The constraint enforcing that only reactions with negative free energies are allowed in the returned pathways (implemented through the constraint in Equation~\eqref{proposed_constraint3}) may be overly stringent.
    Under experimental conditions, the reactions observed to occur do not always align with those predicted to occur theoretically: the thermal decomposition of lithium carbonate is observed to occur in practice despite being predicted to be thermodynamically unfavorable using computed (or experimental) free energies for the reaction.
    However, reactions predicted to be thermodynamically favorable do not always occur in practice because they have a high energy barrier (like the allotropic conversion of diamond into graphite).
    Therefore, there should be some flexibility in delineating the boundary between those reactions that are considered likely and included in the pathway and those that are not, to bring the model closer to reality. 
    Such extensions of the current modeling should be fairly straightforward.
    \item The objective function minimizes the sum of the Gibbs free energies of the reactions in the pathway.
    However, this approach might favor a pathway with more extreme values for the free energies rather than one with more moderate free energy values.
    Therefore, it might be worthwhile to explore alternative objective functions, such as the min-max, in addition to the min-sum used here.
    \item The current formulation disregards the kinetics of individual reactions.
    However, the probability of a reaction occurring is determined to a substantial extent by the reaction barrier height, and not only by the concentration of the reactants and the products.
    As a result, our framework might include a reaction in the returned pathway with a favorable thermodynamic driving force but too high a reaction barrier, making it a slow and rate-limiting step for the entire pathway.
    On the other hand, coming up with a generic `kinetic oracle' with wide applicability to estimate the barrier heights for diverse reactions poses a significant challenge.
    The transition states determining the barrier heights are saddle points on the potential energy surface, unlike reactants or products that are minima and can be located by gradient descent optimization methods. 
\end{enumerate}

Hence, we do not claim this work to be the final in the quest to incorporate thermodynamics into the search for pathways in reaction networks.
Still, we believe that it constitutes a valuable step forward, and we end by summarizing its virtues.
We provide a systematic and thermodynamically informed approach to optimize the computational search for pathways in reaction networks.

Modeling pathways as integer hyperflows gives results that can be interpreted in mechanistic terms,
as each molecule is present in integral quantities.
Formulating the problem as an MILP problem and utilizing an MILP solver allows us to obtain solutions with reasonable computational resources and time.
It allows enumeration of pathways over the integral flow variables, facilitating a systematic exploration of pathways in the reaction network.
Returning several solutions to pathway queries is an asset, as the answer becomes more robust to possible errors in the calculations of chemical potentials of the molecules and to other artifacts of the simplification of reality inherent in any modeling.
Moreover, it offers alternatives to choose from if the optimal solution turns out to be unattainable under controlled industrial reactor or laboratory conditions. 

Finally, the application of our framework to a reaction network, in Section \ref{sec: results}, based on real-life chemistry acts as a proof-of-concept and illustrates the computational feasibility and analytic potential of our methods.

\section*{Acknowledgements}

This work was supported by generous funding from the European Union's Horizon 2021 Research and Innovation program under Marie
Sklodowska-Curie Grant Agreement No.~101072930 (TACsy --- Training Alliance for Computational Systems Chemistry).
The authors would also like to thank Dr.~Maike Bergeler, BASF SE, Ludwigshafen, Germany, for suggesting the use of xTB software.

\begin{scriptsize}
    \bibliography{example_paper}
    \bibliographystyle{icml2025}
\end{scriptsize}


\appendix
\label{appendix}
\section{On assigning the Chemical Potentials}\label{first appendix}

One of the main enhancements to the thermodynamic modeling of reaction networks in this work is a more robust, transparent, and generic method of assigning the chemical potentials under standard physical conditions for molecules, $x^{G^0}_v$,
compared to relying on thermodynamic databases such as (\url{https://pytfa.readthedocs.io/en/latest/thermoDB.html})
using the group contribution method, to infer the change in free energy for reactions, $x^{G^0}_e$, 
as done in~\cite{hatz_therm_python}.
Using quantum chemical methods to assign these chemical potentials quickly becomes computationally resource intensive because the time complexity scales as $\mathcal{O}(N^3)$. The density functional theory (DFT) calculations involve diagonalization of the $N\times N$ Hamiltonian matrix. 
The memory requirement scales as $\mathcal{O}(N^2)$ due to the number of two-electron integrals that contribute to the energy of the system. Moreover, it is an iterative process which is repeated until an acceptable convergence criterion is reached.

Efficient software design necessitates striking a balance among the components in this work: expanding the molecular space, assigning chemical potentials to the molecules in the space, and solving the formulated MILP for queried pathways in the network.   
If DFT calculations are used, the calculation of chemical potentials took a significantly longer time compared to the remaining two components.
Alternative semi-empirical methods offered a faster yet reasonable accurate way to assign chemical potentials to molecules.
Therefore, we used xTB~\cite{grimme_xtb} to estimate the chemical potentials of the molecules in the expanded reaction network.
In the rest of this Section, we describe how we estimate the chemical potentials for the molecules from their representations as graphs.

The molecules in the expanded reaction network were modeled as graphs, where individual vertices represented atoms (labeled by the symbol of the corresponding element), while edges represented bonds. The adjacency matrix of this graph was utilized to determine the connectivity of that molecule. Given a connectivity matrix that satisfies the valences of each atom in the molecule, a three-dimensional embedding for the molecule was generated using VSEPR theory. The \href{https://github.com/openbabel/openbabel/blob/32cf131444c1555c749b356dab44fb9fe275271f/include/openbabel/builder.h#L42}{\texttt{OBBuilder}} class from the Open Babel~\cite{open_babel, openbabel_web} library was used to generate the three dimensional coordinates for the atoms utilizing a combination of VSEPR rules and commonly encountered fragments.

The generated geometry of the molecule was refined using 1000 iterations of conjugate gradient descent or until the change in energy (in atomic units, calculated using the universal force field~\cite{uff}) between successive iterations is less than $10^{-4}$. Next, \href{https://github.com/openbabel/openbabel/blob/32cf131444c1555c749b356dab44fb9fe275271f/src/forcefield.cpp#L1595}{\texttt{OBForceField:: WeightedRotorSearch()}} was utilized to systematically perform a tree search for a low energy conformer of the molecule, by biasing the expansion of the search tree toward the current low energy conformer. At each iterative expansion step, 25 conformers were generated, and 100 conjugate descent steps are executed on each. This leads to the generation of conformers with successively lower energies over subsequent iterations. The conformer with the lowest energy identified at the end of this search process was further subjected to 250 iterations of conjugate descent or until the energy tolerance of $10^{-5}$ was reached. The energy of a conformer was estimated at each step using the universal force field~\cite{uff}. The result of this search process, the configuration of the low-energy conformer, which contains the three dimensional Cartesian coordinate of each atom in the molecule, was written into a \texttt{.xyz} file. 

The \texttt{.xyz} coordinate file generated by OpenBabel serves as an input file for xTB.
xTB performs an optimization of the geometry of the structure, employing an energy tolerance of $5\times 10^{-8}$ Eh and a gradient tolerance of $5\times 10^{-5}$ Eh bohr$^{-1}$. This is followed by the calculation of the Hessian matrix for the optimized structure and the chemical potentials at $298.15$ K using standard translational, rotational, vibrational, and electronic partition functions from statistical mechanics under the coupled rigid-rotor and harmonic oscillator approximations.

\section{On the range of the concentration variable}\label{second appendix}

In addition to the free energies for the molecules $x^{G^0}_v$, assigned by the thermodynamic oracle, the limits on the log concentration values of the molecules $\left[x^K_v\right]_{\text{min}}$ and $\left[x^K_v\right]_{\text{max}}$ have to be assigned by the user.
The upper limit for the concentrations is a physical limit bounded by the concentration of the pure substance. Concentration is expressed in units of moles per liter (mol L$^{-1}$) for liquids and solutions, while in terms of partial pressures (in Pa) for gases. For liquids, the maximum possible concentration value of the reactant would be that of the pure liquid, which is physically constrained by the following:
\begin{align*}
    1 \text{ mol of pure liquid} &\equiv M \text{ kg of pure liquid} \\
    1 \text{ mol of pure liquid} &\equiv \frac{M}{\rho} \text{ L of pure liquid} \\
    \text{Maximum concentration} &\equiv \frac{\rho}{M} \text{mol L$^{-1}$}
\end{align*}
For gases, the maximum concentration is imposed by the physical constraint that it has to remain (an ideal) gas following the equation of state $PV = nRT$ so that the equations from the kinetic theory of gases might still be applied to it. The concentration of a gas under some given physical condition of pressure $P$ and temperature $T$ is
\begin{equation*}
    \frac{n}{V} = \frac{P}{RT}
\end{equation*}
As the pressure is increased, or the temperature is decreased to increase the concentration, the gas ceases to behave as an ideal gas and starts to exhibit properties of the vapor phase. A change in the physical state of the substance would have a free energy of phase reaction involved, and it can no longer be modeled only using the equations considered here. 

The lower limit of the reactants is imposed by practical constraints because there are limits to the dilution that can be achieved under laboratory conditions. For a solution, an infinite dilution would lead to no solute (reactant) being present, and hence the reaction would not occur. For gases, there are lower limits to the pressures that can be achieved in laboratory vacuum settings. Reactions occurring in the solid phase (or on surfaces) are governed by different physical equations and cannot be expressed by this framework. For all practical purposes, we can assume that the reactants are solutions or gases satisfying the condition of a well-stirred reactor.

For this work, we have chosen the limits in the concentration variables as $\left[x^K_v\right]_{\text{min}} = -3$ and $\left[x^K_v\right]_{\text{max}} = 1$. These correspond to physical values of the concentrations between $10^{-3}$ mol lit$^{-1}$ and $10^{1}$ mol lit$^{-1}$. Although minimolar concentrations can be achieved under laboratory conditions using standard fludic procedures, a concentration of tens of molars is often encountered for pure liquids (the concentration of pure water is $55.56$ mol lit$^{-1}$). Although the same ranges for the log concentration values are used here for all of the molecules in the reaction network, the user might choose to assign different ranges for different molecules to make the model more realistic.

\section{Graph Transformation rules used}\label{third appendix}
The graph transformation rules used and the expanded reaction network are listed in this appendix.
\newcommand\ruleScale{0.8}

\begin{figure}[!h]
    \begin{subfigure}{0.5\textwidth}
        \centering
        \dpoRule[scale=\ruleScale]{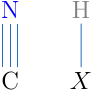}{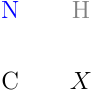}{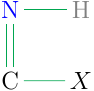}
        \caption{Rule 1: Addition of water or ammonia to hydrogen cyanide (nitrile group) where $X \in \{\mathrm{N}, \mathrm{O}\}$ to form an imine.}
        \label{fig:rule1}
    \end{subfigure}
    \begin{subfigure}{0.5\textwidth}
        \centering
        \dpoRule[scale=\ruleScale]{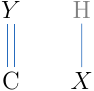}{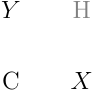}{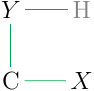}
	    \caption{Rule 2: Addition of water or ammonia to a double bond where $X, Y \in \{\mathrm{N}, \mathrm{O}\}$ (in an imine or carbonyl group respectively).}
        \label{fig:rule2}
    \end{subfigure}
    \begin{subfigure}{0.5\textwidth}
        \centering
        \dpoRule[scale=\ruleScale]{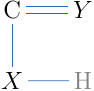}{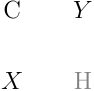}{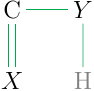}
	    \caption{Rule 3: 1,2 to 2,3 bond shift of the double bond to form tautomers where $X, Y \in \{\mathrm{N}, \mathrm{O}\}$.}
        \label{fig:rule3}
    \end{subfigure}
    \begin{subfigure}{0.5\textwidth}
        \centering
        \dpoRule[scale=\ruleScale]{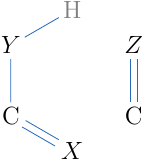}{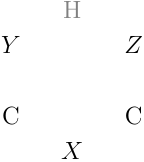}{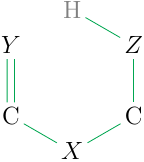}
	    \caption{Rule 4: Addition of two molecules involving six atoms, where $X, Y, Z \in \{\mathrm{N}, \mathrm{O}\}$, modeling an aldol like addition to carbonyls and imines. }
        \label{fig:rule4}
    \end{subfigure}    
    \caption{Graph transformation rules (or reaction templates) used to generate the reaction network \ref{fig: hypergraph_for_RN} in which pathways were queried. The reverse of rules \ref{fig:rule2}, \ref{fig:rule3} and \ref{fig:rule4} was used, while the reverse of rule \ref{fig:rule1} was \emph{not} used while expanding the hypergraph.}
\end{figure}

\begin{figure}[thb]
    \centering
    \includegraphics[width=0.5\textwidth]{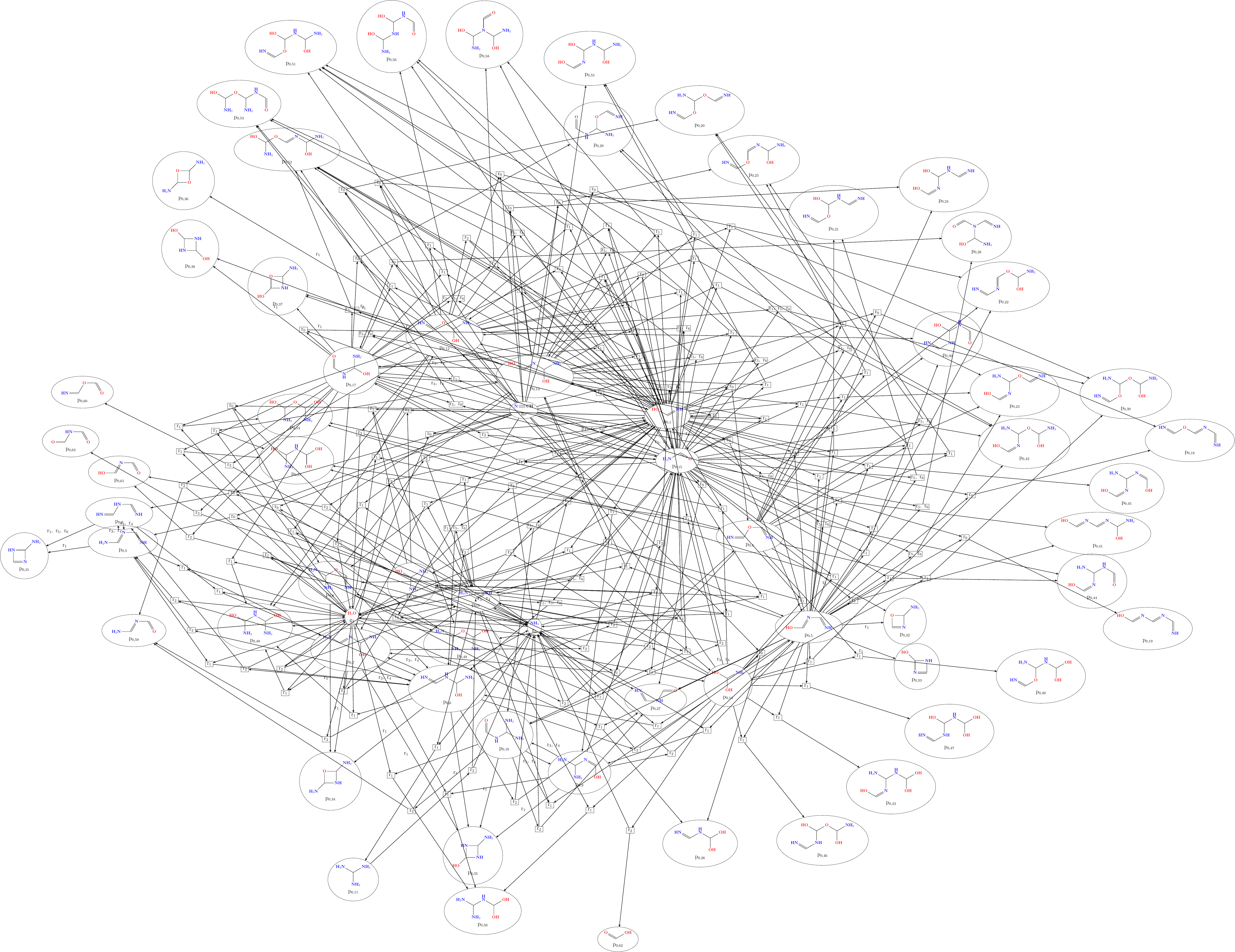}
    \caption{Hypergraph with $|V| = 67$ and $|E| = 202$, representing the reaction network considered in this work, provided here for the sake of completeness.}
    \label{fig: hypergraph_for_RN}
\end{figure}

\end{document}